\documentstyle[12pt]{article}
 \if@twoside
 \else
 \fi

 \parskip \medskipamount
\begin{document}
\hfill{UMDHEP PP93-170}
\vspace{24pt}
\begin{center}
{\large\sc{\bf Amplitude Zeroes in Collinear Processes or What Is
Left from a Factorizable 2d Model in Higher Dimensions.}}
\baselineskip=12pt
\vspace{35pt}

G. Gat$^1$ and B.Rosenstein$^2$
\vspace{24pt}

$^1$  Physics Department\\
University of Maryland \\
College Park, MD 20742\\
U.S.A \\
\vspace{12pt}
$^2$  Institute of Physics \\
Academia Sinica\\
Taipei, 11529\\
Taiwan\\
\vspace{60pt}
\end{center}
\baselineskip=24pt
\begin{center}
{\bf abstract}
\end{center}

We show that for collinear processes, i.e. processes where the incoming and
outgoing momenta are aligned along the same line, the S-matrix of the
tree level 2+1 dimensional Thirring model factorizes: any S - matrix element is
a product of $2\rightarrow 2$ elements.
 In particular this means nullification of all collinear $2 \rightarrow n$
amplitudes
 for $n > 2$.

\vspace{100pt}

\vspace{30pt}

\pagebreak

The behavior of amplitudes for production of a large number of particles
has recently been studied  within the framework of perturbation theory.
The tree and one loop level contribution to the $2 \rightarrow n$ amplitudes
at threshold (i.e. all produced particles are at rest) has been calculated
exactly for $\phi^4$ scalar theory (broken and unbroken phase) using various
methods \cite{volosh,pap}. The
phenomenon of ${\it nullification}$ was discovered: for all $n \ge n_0$
the amplitude
${\cal A}(2 \rightarrow n)$ vanishes when all n particles are produced
at rest (threshold). In the unbroken phase $n_0=4$,
while in the broken phase $n_0=2$.
The phenomenon is rather exclusive and occurs only in
special models \cite{brown}.
For example a general scalar model with just two fields doesn't exhibit
nullification.

Argyres,
Kleiss and Papadopoulous \cite{pap} proposed a general method based on
constructing generating
function for the amplitude $A_{1 \rightarrow n}$. The recursion
relations for it
reduces to a second order differential equation with a potential which
is (in all known cases) one of the solvable
reflectionless potentials of
quantum mechanics (quantum field theory in $d=1$).
They pointed out that
the study of the relation between nullification at kinematical threshold and
integrability in $d=2$ space time dimensions could provide us with a new
insight into multiboson production processes.

In this note we consider a generalization of the above results in the following
direction.
Instead of considering all the particles at threshold i.e. with
zero momentum we only restrict them to move along the same line.
We consider the simplest\footnote{from the
point of view of performing perturbation theory} 2d - integrable model , the
Thirring model in higher dimensions. First we show that the 2d type
factorization which
immediately implies nullification of all $2 \rightarrow n$ collinear
amplitudes for
$ n \ge 2$ is preserved on tree level 3d model for collinear
processes despite nonelasticity. The phenomenon seems to
be even more exclusive then the nullification. For example, (a nonsolvable in
2d) model like $\phi^4$ does not possesses such a property .

 The massive Thirring model \footnote{The  $d>2$ dimensional Thirring model is
 perturbatively nonrenormalizable (however it can be shown \cite{gomes}
 that non-perturbatively in 3d four - Fermi interactions are
in fact renormalizable). Note that the nullification
occurs in nonrenormalizable models: $\phi^4$ is nonrenormalizable for $d>4$.
There is no direct relation between the nullification and the ultraviolet
properties so one can gain information on the behavior of
 multiparticle amplitudes by considering the first few orders in perturbation
 theory even in apparently nonrenormalizable theories.}
 \begin{equation}
 {\cal L}= \bar \psi( i\partial \! \!\! /-m)\psi +\frac{g}{2}(\bar \psi
 \gamma_{\mu} \psi )^2
 \end{equation}
 belongs to the special class of non-trivial 1+1 dimensional models
 with $n\rightarrow m$
S - matrix elements factorizing into a product of $2\rightarrow 2$ elements,
 which in turn
are purely elastic (other solvable models are
nonlinear $\sigma$ model, Gross - Neveu, ... \cite{zz}). The factorizability
 is a consequence of the
 infinite number of conservation laws
 that exist in these models. Together with the general properties of
 analyticity, crossing and unitarity of the S - matrix, it
enables the exact
 determination of their S matrix.

In higher dimensions, purely elastic $2 \rightarrow 2$
nontrivial S-matrix
 necessarily contradicts relativistic invariance or
basic analyticity properties of the S-matrix \cite{Col}.
Also the factorization like that in 2d is not possible.
 However like in 2d one can get some hints about the structure of these
theories and
possible factorization studying the perturbation series.
In two dimensions one is able to spot the factorization
within perturbation series using so called "cutting rules"
\cite{kallen} It is interesting that there are cutting rules also in higher
dimensions. A cutting rule states that any one loop
 diagram with n external legs in space time of dimension d can be reduced
 to products of tree level diagrams and a one loop diagram with d external
 lines. This feature gives rise to the hope that for some models even
 in $d > 2$ the $S$ matrix can be built up (algebraically) from a finite
 number of independent functions \cite{berg1}. In particular in three
dimensions
 the only irreducible diagrams are the two ("fish") and three ("triangle")
legged
 one loop diagrams along with the tree level diagrams. Thus,  it is
possible in principle that any S-matrix element
 of some model might be algebraically built up from
  $2 \rightarrow 2 $, $2\rightarrow 4$
 and $3 \rightarrow 3$ amplitudes.

We start by showing the complete factorizability of the scattering amplitude
for collinear processes at tree level. The first step in the induction is
an explicit calculation of the $3 \rightarrow 3$ amplitude for such processes.
   The $3 \rightarrow 3 $ amplitude at tree level is (see fig.1):
\begin{equation}
{\cal A}_{33}=\frac{3g^2}{4}\sum_{\pi(p_1,\ldots p_6)} \bar{u}(p_{\pi_1})
\gamma_{\mu}u(p_{\pi_4})
\bar{u}(p_{\pi_2})\gamma^{\mu} \frac{i}{p_{\pi_1} \! \! \!\! \! \! \!/
+p_{\pi_2} \!\! \! \!\! \! \!/-p_{\pi_4} \! \!\!\!\!\!\!/-m}
\gamma_{\nu}u(p_{\pi_5})\bar{u}(p_{\pi_3})\gamma^{\nu} u(p_{\pi_6})
\end{equation}
 one thus has to show that
 for all values of collinear  external momenta for which the propagator
 (see fig.1) is
 off shell the amplitude vanishes.
 Using the following decomposition of the propagator:
\begin{equation}
\frac{i}{p \! \! \! /-m}={\cal P} \frac{i}{p \! \! \!/-m }+\delta (p^2-m^2)
\end{equation}
(where ${\cal P}$ means principle value part) it is possible explicitly
see that the principle
value part of the amplitude vanishes for arbitrary collinear momenta.
Note that this cancellation is non trivial
because the restriction to collinear processes is by
no means equivalent to the 1+1
dimensional interaction. The third component of the current $J_{\mu}=
\bar \psi \gamma_{\mu} \psi$ gives a nonvanishing contribution to the
 amplitude. We have also checked the $ 3 \rightarrow 3 $ and $ 4 \rightarrow
 4$ amplitudes in the 3+1 dimensional Thirring model. The principle value
 part of these amplitudes does not vanish even for collinear processes in
 this case.

The next step is to  use induction  to prove that all
$n \rightarrow
n$ amplitudes factorize. We thus assume that ${\cal P}{\cal A}_{nn}^{tr}=0$
for all $n \le N-1$
and prove  ${\cal P} {\cal A}_{NN}=0$ . We shall only outline the proof
here since essentially it is the same as Berg's \cite{berg2} for the 1+1
dimensional model. One fixes $N-2$ of the momenta and
considers the amplitude ${\cal A}_{NN}$ as a function of the remaining
 independent
variables, say ${\bf p_1}$. We let ${\bf p_1}$ take complex values and denote
${\cal A}_{NN}=w(z)$. If the amplitude is not constant it satisfies an
irreducible algebraic equation of the form
\begin{equation}
\sum_{j,k=0}^{r,s} a_{ij}w^{j}z^{k}=0
\end{equation}
where $r \ge 1$ and $a_{kk} \ne 0$ for at least one $k$. This equation defines
a compact Riemann surface ${\cal F}$ \cite{kra}. For each point on this
surface
one can choose local coordinates  such that $w$ is a single valued function
on it. On the whole surface the number of poles and zeroes of $w$ is equal.
 Using the induction hypothesis it is now possible to show that the maximum
degree of divergence of ${\cal P} {\cal A}_{NN}$ is $ \sim z^2$ for $z
\rightarrow \infty $ . By Fermi statistics $w$ has at least $N-1$ zeroes and
therefore
${\cal P}{\cal A}_{NN}=0$ for $ N \ge 3$.
As a check we also calculated the $4 \rightarrow 4$ amplitude for various
collinear momenta and verified that the principle part of the amplitude
vanishes.

The factorization of the S-matrix for collinear processes automaticaly
  forbids particle
production. This is because any such processes is related by crossing to an
$n \rightarrow n$ amplitude where one of the intermediate propagators is
off-shell. In particular threshold production i.e. $2 \rightarrow n$ where
$ n > 2$ and the produced particles have spatial momentum zero is a special
case of a collinear process. It is clear from the previous discussion that
the amplitude for such processes in the Thirring model vanishes. We have also
checked what happens when one of the particles momenta is off the line.
In this case the principle value of the amplitude deviates from zero. This
means that simple factorization to $3 \rightarrow 3$ and $2 \rightarrow 2$
amplitudes does not occur  in this model.  A similar
calculation for the 3+1 dimensional Thirring model
shows
no sign of this property (i.e even for collinear processes ${\cal P}
{\cal A}_{NN} \ne 0$).

In the two dimensions one can extend this proof to all orders in  loop
expansion by using a cutting rule due to K\"{a}llen and Toll \cite{kallen}.
This cutting
rule states that any $n$-legged one loop boson \footnote{since any fermionic
one loop
diagram can be written as a sum over bosonic diagrams and their derivatives
with respect to external momenta this is true also for such diagrams}
 diagram (see fig. 2) can be written
as a product of tree level graphs where each term in the sum is multiplied
by a one loop integral with two external legs and momentum assignment
determined
by the cut. It is clearly seen then that if tree level factorization has
been proven no particle production is possible through any one loop diagram.
The K\"{a}llen-Toll cutting rule has obvious generalizations to higher
dimensions. For an $m$ legged loop diagram ($ m > d$) in $d=3$ it reads (see
fig. 2):
\begin{equation}
\int \frac{d^3 k}{(2 \pi)^3} \prod_{j=1}^{m} \frac{1}{(k-p_j)^2-m^2+i
\epsilon} =\frac{1}{2} \sum_p \sum_{s=\pm} \left[ {\cal T}_p^s \int
 \frac{1}{(2 \pi)^3} \prod_{i \in P} \frac{1}{(k-p_i)^2-m^2+i\epsilon} \right]
\end{equation}
where the sum is over all partitions $P= \{ \{ \pi_1, \pi_2 \} \}$
with $ \{ \pi_1, \pi_2  \} \in \{ 1,\ldots m \}$ and
\[ {\cal T}_p^s= \prod_{r \in P} \frac{1}{(k_p^s-p_r)^2-m^2}  \]
where $k_{ij}^s$ are the two solutions of $(k-p_i)^2=m^2$ $(i \in P)$.
 any one loop diagram can thus be written as a
sum over tree level diagrams multiplied by a one loop diagram with three
external lines with momentum assignment determined by the cuts. We therefore
see that since triangle diagram necessarily allows particle production,
non particle number preserving processes are inevitable,
in agreement with the general theorems \cite{Col}.

To conclude, we considered generalizations of some
integrable models to higher
dimensions. It was shown that  for the 3d massive Thirring model collinear
factorization survives at the tree level.
 The nullification of the threshold $2 \rightarrow n$
amplitudes is an immediate consequence of this property and follows from
the collinear factorizability property.
The assumption of simple inelastic factorization as implied by the
generalization of the cutting
rule (e.g. in $d=3$ writing ${\cal A}_{4 \rightarrow 4} $ as a product
of ${\cal A}_{2 \rightarrow 2} $ and ${\cal A}_{3 \rightarrow 3}$ ), proved
to fail for the Thirring model. However, it would be interesting if such a
model could be found.

\pagebreak

\begin{figure}
\vspace{2in}
\caption{The $3 \rightarrow 3$ amplitude. The wavy line represents a
 fictitious photon}
\end{figure}

\begin{figure}
\vspace{2in}
\caption{The Kallen-Toll cutting rule. A reduction of the triangle diagram in
1+1 d . B reduction of the box diagram in 2+1 d.}
\end{figure}
 \end{document}